# Design Exploration of Lightweight Interactions for Awareness-Supporting Technologies in Hybrid Work

Liu, Lu*[a]; van Essen, Harm[a]; Eggen, Berry[a]

[a] Eindhoven University of Technology, Eindhoven, The Netherlands
* l.liu3@tue.nl

Hybrid work settings often lack the informal communication that naturally emerges from spontaneous encounters and ambient awareness of coworkers' activities, potentially hindering team collaboration. To address this challenge, we explored how lightweight interactions can be integrated into awareness-supporting technologies for fostering informal communication. Our experiential design approach focused on how information is perceived and processed rather than explicit content exchange. Through brainstorming, speculating, and prototyping, we explored the design space for small hybrid teams. By annotating and analyzing design concepts, speculative scenarios, and prototypes, we developed a framework that identified design options for lightweight interactions and methods for integrating them with information displays.

## 1   Introduction

In the post-COVID-19 era, many people have adopted the hybrid work style, flexibly working remotely and onsite (Teevan et al., 2022). Despite advances in online work technologies, the hybrid work model has presented communication challenges including discrepancies between virtual and real-world information (Mahajan et al., 2024), extra articulation work before online communication (Bjørn et al., 2014; Busboom & Boulus-Rødje, 2023), and fatigue from excessive videoconferencing (Bailenson, 2021; Nesher Shoshan & Wehrt, 2022). These issues hindered informal communication that naturally occurs in onsite settings. For instance, initiating face-to-face conversations in an office is easy, but in remote or hybrid settings, people may find it difficult to gauge colleagues' availability, coordinate meeting times, or even feel reluctant to engage in video calls. As prior research suggested, informal communication contributes to team collaboration, cohesion, and creativity (Kraut et al., 1990; Lederman, 2019; Whittaker et al., 1994). Therefore, we intend to design technologies that promote informal communication among hybrid workers.

Rather than focusing on explicit content exchange tools, which are already widely available, we turn to awareness-supporting technologies. Extensive research has pointed out the importance of maintaining social awareness of individual and team activities for active informal communication (Dourish & Bly, 1992; Gutwin & Greenberg, 2002; Streitz Norbert and Prante, 2007). These technologies provide coworkers' activity information, typically using abstract visualizations, ambient displays, or tangible interfaces to convey contextual information without explicit messaging.

Recent research into remote and hybrid work experiences indicates that the difference lies not in the amount of information but in how it is received compared to onsite work (Lal et al., 2023; Liu et al., 2022; Nesher Shoshan & Wehrt, 2022). For example, in a physical office, one might overhear colleagues having a coffee break, whereas checking status icons for such information from online systems feels unnatural and often inaccurate.

These subtle yet impactful differences in how information is perceived and processed highlight design opportunities for experience-oriented technologies. Our previous research (Liu et al., 2024) has identified experiential goals for awareness-supporting technologies in hybrid work, including enabling users to perceive information easily, feel others' social presence, make sense of information in context, and exchange simple responses.

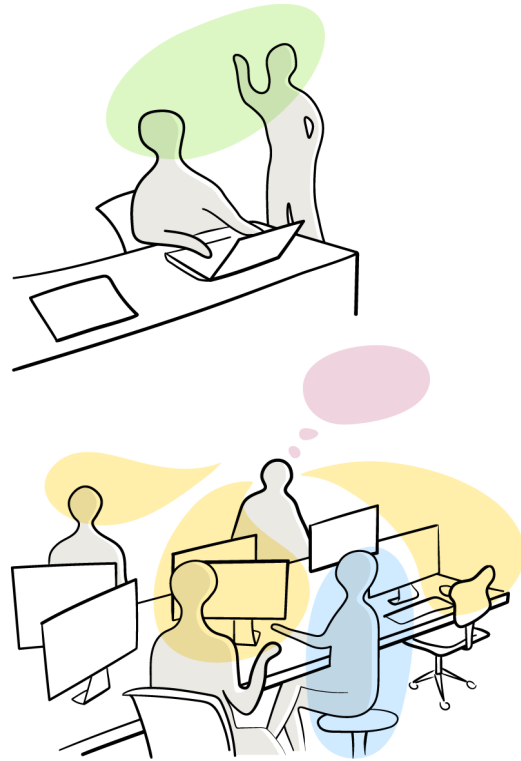

Figure 1. An illustration of the experiences that we aim to achieve in our information technology designs

While this prior research explored a rich design space for information displays, the design aspect of built-in lightweight interactions remains comparatively underdeveloped. Lightweight interactions enable expression without explicit content exchange. Tangible interfaces commonly feature these interactions by supporting easy user input. Examples include 'Nimio,' a desk toy that detects movement (Brewer et al., 2007), and 'Mood Squeezer,' which translated squeezing softballs into visualized emotions (Gallacher et al., 2015). Interactive environments further illustrate their usage, particularly through embodied interactions. For example, 'Traces,' an in-situ floor display, visualized walking trajectories while inviting expressive bodily movement such as drawing shapes (Monastero & McGookin, 2018). Similarly, ambient awareness systems can also benefit from lightweight interactions, as seen with 'SnowGlobe,' a lamp conveying human presence between two remote rooms, where shaking the local lamp triggered blinking of the remote lamp (Visser et al., 2011).

Despite these examples demonstrating the potential of lightweight interactions in awareness-supporting technologies, a systematic approach to their design and integration remains lacking. This gap motivated the exploration of designing and integrating lightweight interactions into awareness-supporting technologies that presented in this pictorial. To achieve this, we adopted the Research-through-Design (RtD) approach (Zimmerman et al., 2007), using design concepts, speculative scenarios, and prototypes as inquiry materials. Specifically, we focused on small, collaborative teams, as their work heavily relies on enhanced social awareness and informal communication.

## 2   Overview of the RtD process

This RtD process comprised iterative cycles of ideation, speculation, and prototyping, during which we documented the process and design knowledge about lightweight interactions.

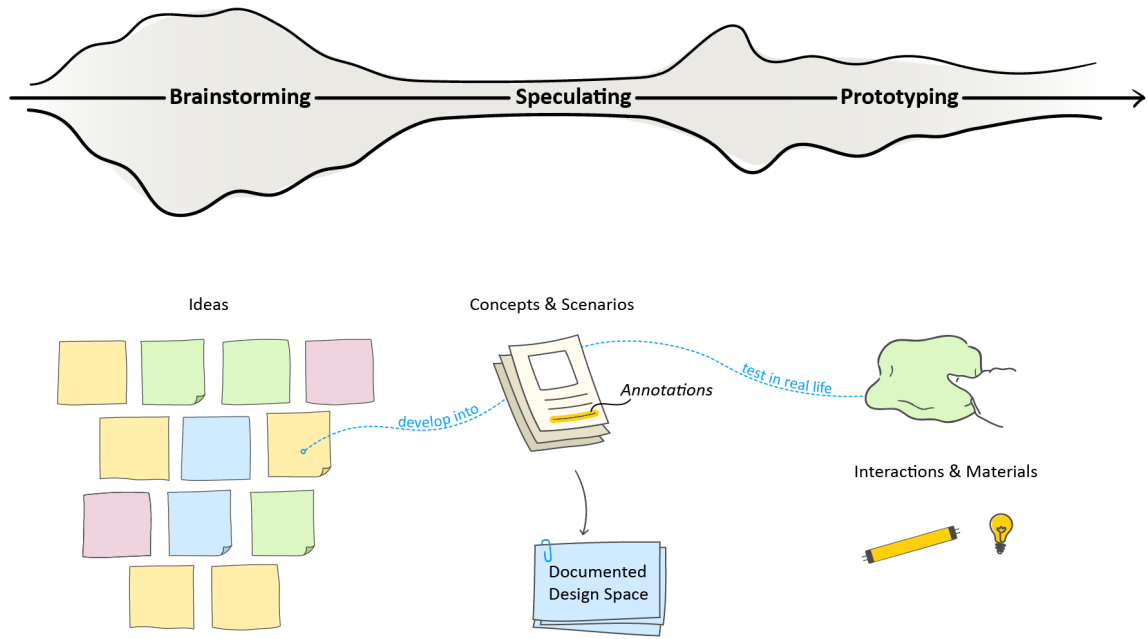

*Figure 2. An overview of our research-through-design process, comprising three stages: brainstorming, speculating and prototyping, with their respective design activities and generated materials*

We began with **brainstorming** sessions involving the main researcher and two external designers. We provided a design context to scope the challenge and the existing design space to facilitate ideation. From the generated ideas, we identified promising design directions based on their potential for supporting awareness and lightweight interactions.

These initial directions informed our selection of three concepts featuring distinct displays and interactions. We developed each into a complete design concept by specifying their technical components and interaction mechanisms. To envision their usage in hybrid work settings, we created speculative scenarios (Auger, 2013). While **speculating** these scenarios, we drew on the notion of 'future mundane (Crabtree et al., 2025),' depicting familiar work scenes and events to surface the experiences people might have with our awareness technologies. Through annotation and analysis of the concept and scenarios (Vasconcelos et al., 2024), we identified recurring patterns in user interactions and intentions. This analysis revealed three design principles for lightweight interactions and their integration methods, which we then synthesized into a design space framework.

Guided by the design space, we used accessible materials for rapid **prototyping** to transform the concepts into tangible experiences. Through iterative making and testing, we documented how theoretical principles translated into material constraints while material affordances revealed unanticipated design opportunities. This hands-on exploration surfaced practical considerations that refined our understanding of implementing lightweight interactions in real contexts (Mäkelä, 2007).

# 3 Brainstorming Design Ideas

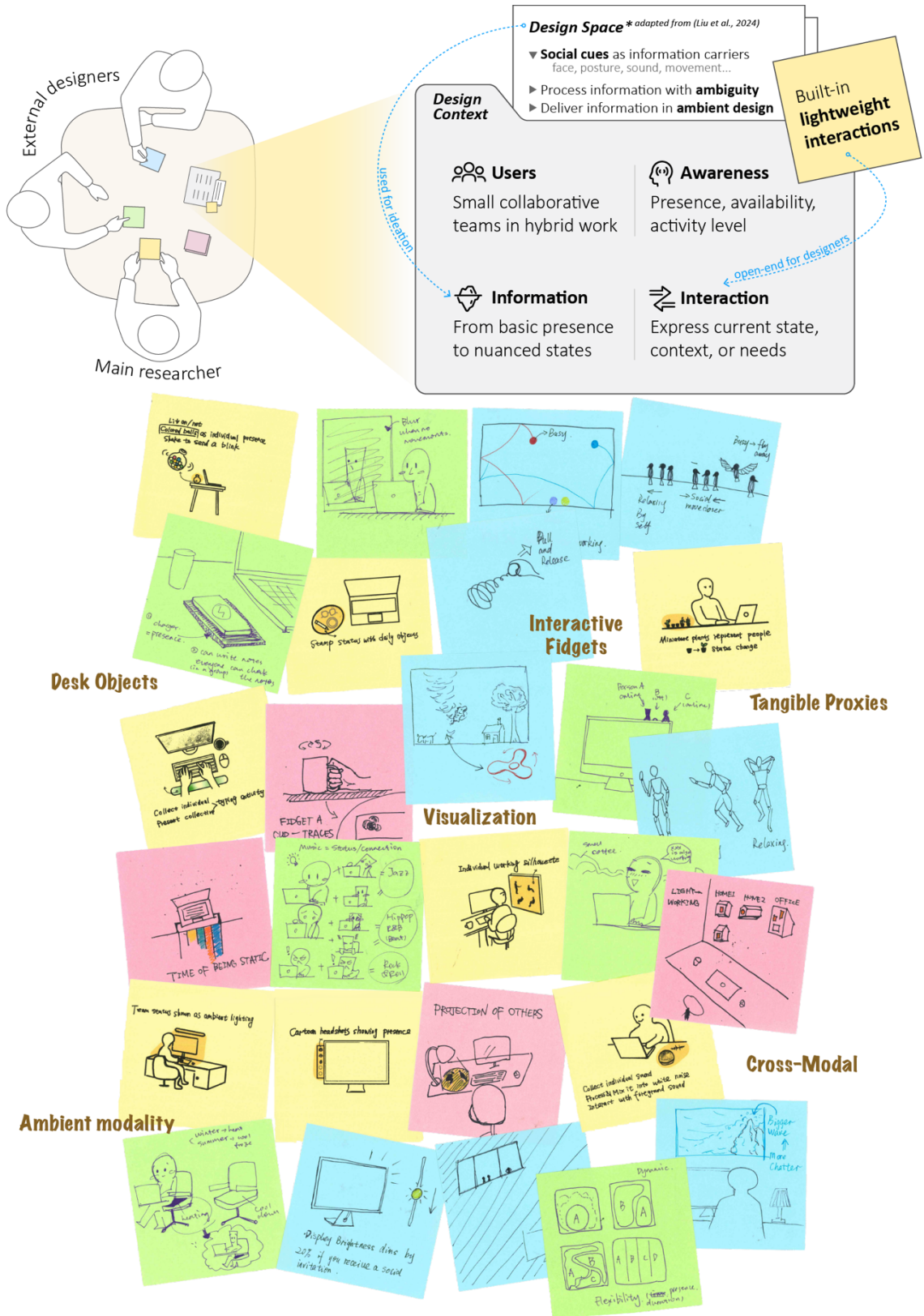

Figure 3. A visualization of the brainstorming process, showing the materials used and the design ideas generated

# 4 Speculating Usage Scenarios

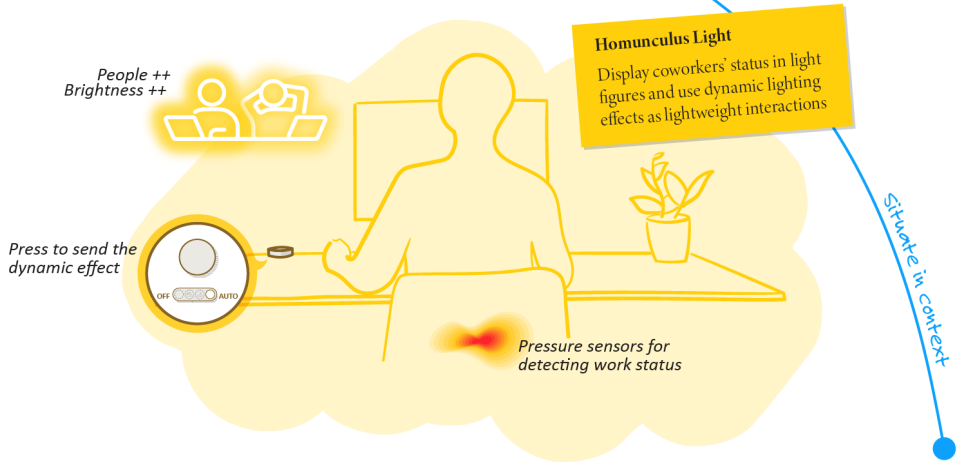

*Figure 4. A visualization of the speculating process, using the concept Homunculus Light to illustrate the three steps*

**Cup Imprints**
Display cup stains as coworkers' status and use simple sketching as lightweight interactions

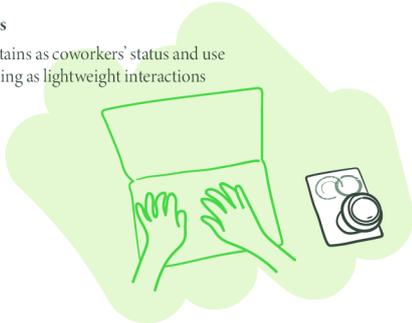

**Self-created expression built aside displays**

sketching    aside the cup stains

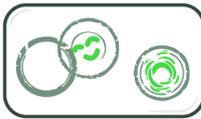

After working at his desk for a while, Max wants to take a short break. As he reaches for his coffee on the desk, he glances at the mat display under the cup. The mat shows "stains" from his teammates' cups, allowing him to gauge their status. As Max lifts his cup, a swirl pattern emerges in his own cup stain, signaling a status update. Looking at the cup stains, Max knows that his two teammates are working today as well. A hand-drawn smile appears in a round cup stain, while Max is sipping his coffee and looking at the mat, which also brings a smile to Max's face. It feels like someone is encouraging him from afar. Concluding his brief coffee break, Max feels revitalized.

input with embodied interactions
send status change information
make sense in context

Lightweight interactions can be **interpreted**

*What makes interactions lightweight?*

*Why use lightweight interactions?*

feeling social presence
receive opportunistic social feedback

Use lightweight interactions to **signal social presence and rapport**

**Preset expression built apart from information displays**

3 types of sounds    sound vs light

*How to design lightweight interactions?*

Use lightweight interactions to **convey communicational intentions**

check information intentionally
send intentional social quest

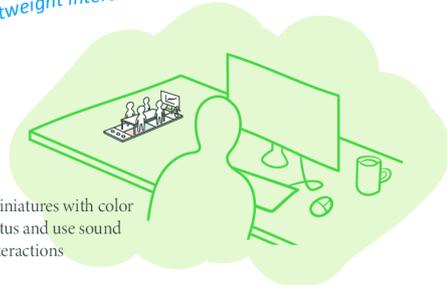

**Avatar Island**
Display coworkers as miniatures with color light indicating their status and use sound effects as lightweight interactions

Sam is working in the office today, but the space feels unusually quiet—most of the desks around are empty. On Sam's desk, a miniature of the office displays avatars representing her teammates. The colors of the square "floor" beneath each avatar indicate their working status. Throughout the morning, Sam notices that the white-lit floors appear one by one, confirming their suspicion: most teammates have opted to work from home today. As Sam focuses on working, a notification pops up about an upcoming seminar. Sam hesitates, unsure whether to attend, and is curious if any teammates plan to join. Hoping to gauge their interest, Sam selects the gentle bubbling sound from the miniature interface and sends it out to the team. The sound carries her subtle question, inviting a response without disrupting anyone's focus. Sam waits, and the silence lingers. They think that maybe everyone is too busy and the seminar is not that important. Deciding to skip it, Sam goes back to their work.

Lightweight interactions can be **ignored**

transient interactions
ease with no responses

*Figure 5. A visualization of concept Cup Imprints and Avatar Island, focusing on how we annotated and analyzed the concepts and scenarios and documented design knowledge*

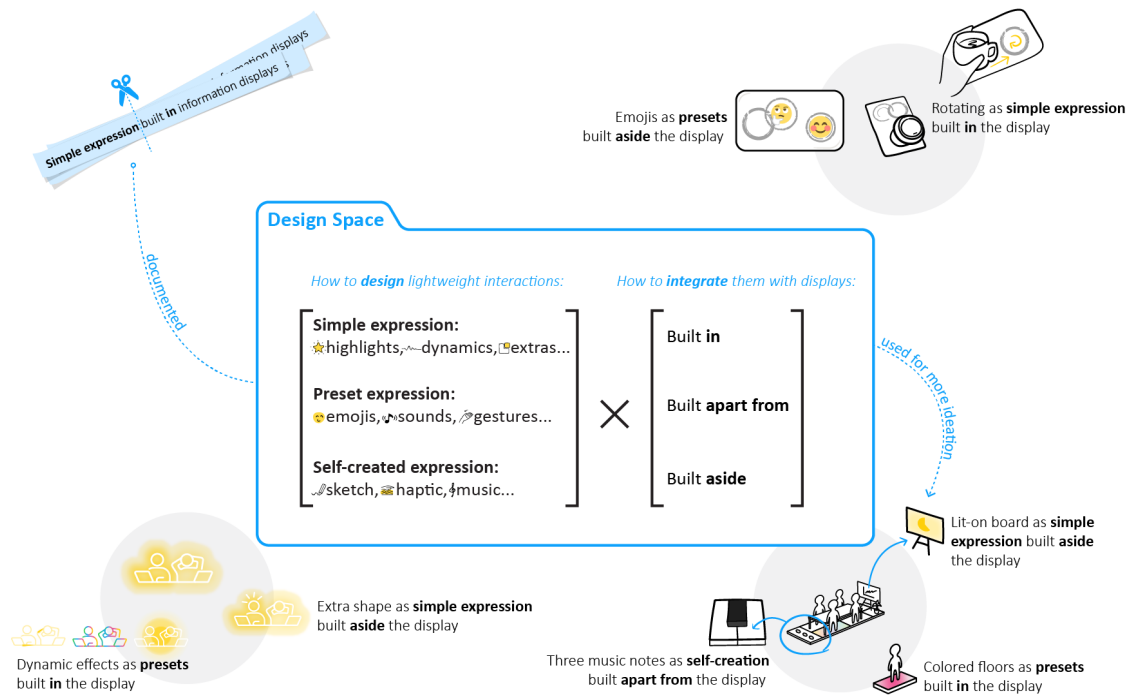

*Figure 6. A visualization of the documented design space for lightweight interactions, along with its exemplar usage in generating more varied interactions for each concept*

While annotating, we asked three questions: How are lightweight interactions designed within each concept? Why do hybrid workers use lightweight interactions? What distinguishes lightweight interactions from content-oriented tools in hybrid work communication? Through this analysis, we identified key insights illustrated as sticky notes in *Figure 4–Figure 5*.

Design Space (blue notes): By examining interactions within the concepts, we annotated three ways of designing lightweight interactions. By reorganizing the annotations, we categorized them into two groups: design options for lightweight interactions and methods for integrating them with information displays. Combining these options and integration methods flexibly allowed us to generate more variations for each concept (see in *Figure 6*). In this way, we provided a generative framework for creating lightweight interactions in awareness-supporting technologies.

User Motivations (yellow notes): By examining user intentions and experiences in the scenarios, we identified three reasons hybrid workers use lightweight interactions. First, they signal social presence and rapport without explicit communication, resembling nonverbal cues like a smile or nod. Second, they enable instant feedback during shared social moments like morning greetings or pre-meeting coordination. These 'human moments' (Hallowell, 1999) can foster social awareness and connections, which are often diminished in hybrid work settings (Fayard et al., 2021). Third, they convey spontaneous communicational intentions, sparking unplanned conversations.

Interaction Characteristics (pink notes): These interactions are inherently interpretable because their meanings emerge from context rather than explicit labels. When integrated with displays, they subtly draw attention without disruption. Their implicit and transient nature allows them to be easily ignored, contributing to their lightweight quality.

# 5 Rapid Prototyping & Testing

Throughout prototyping, we documented emerging design knowledge from making things and rapid testing with the authors' coworkers. Three types of insights emerged: (1) design requirements for different types of lightweight interactions, (2) how information displays can manipulate anonymity in group systems, and (3) practical implementation considerations for each concept.

## 5.1 Nudging for engagement

We created functional prototypes using affordable smart lighting devices. We cut tracing paper into human figure covers for smart light bulbs and shaped light strips into one-line art. Both prototypes provided aesthetics for home office environments. Despite their abstract nature, dynamic lighting effects nudged users toward engagement by inviting interpretation and offering easy input.

For the intended information, paper-covered bulbs provided clearer presence indication. The brighter the paper figure, the more presence was there. However, the light strips' scattered brightness made status differentiation challenging.

For the interaction testing, dynamic changes were easily perceived from the environment. The light strips provided a nuanced experience, distinguishing people's status changes (slow and single-direction flows) and expressive signals (fast and bi-directional dynamics). For light bulbs, a blinking effect is hardly distinguished from a flaming one during focused work.

In this concept, while individuals hide behind the collective lighting effects, the rich color and animation capabilities offered potential for individual expression within the group display.

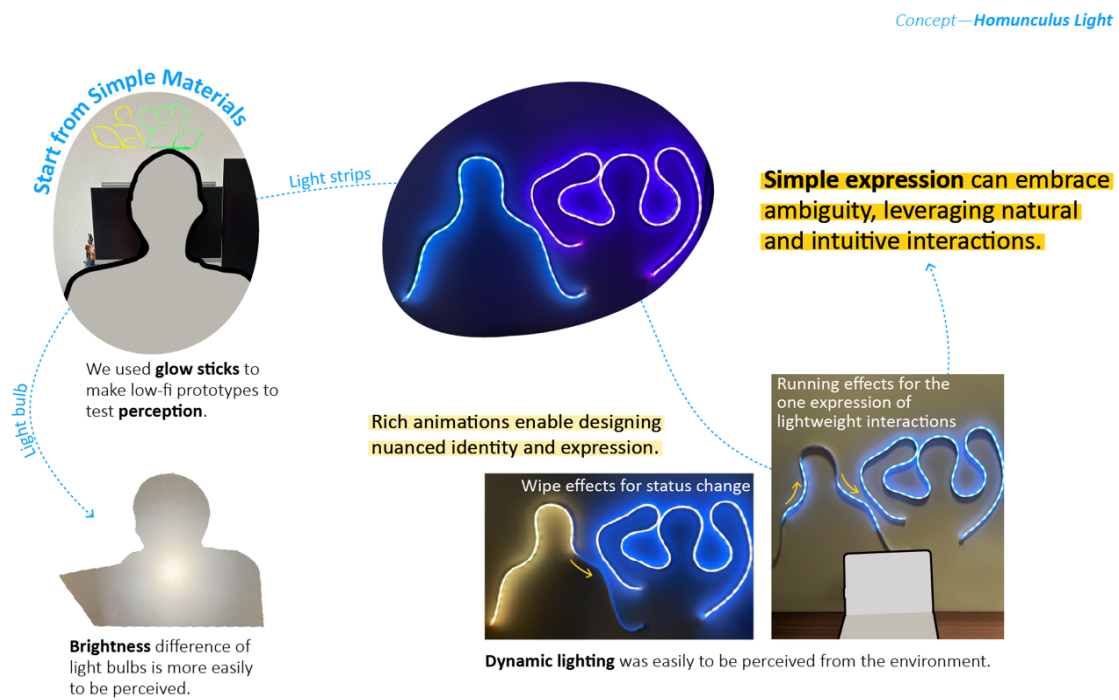

*Figure 7. A visual documentation of prototyping the concept Homunculus Light*

## 5.2 Constraining for creativity

This concept leveraged people's routine activities in the workplace, specifically individual interactions with drinking cups. We began by observing the 'collections' of cups in workplace settings, which typically included both universal cups provided by employers and unique cups brought by employees. This insight informed a semi-anonymous system where users could blend in or stand out through cup choice. Cup molding details offered a rich visual language for identity expression, with repeated use enabling recognition over time.

The imprints left by cups created a visual space for drawing to convey intentions and reactions. Whether stamped on clay or stained on a cork mat, people made only simple doodles due to thick strokes, limited canvas, and clumsy fingers. While on a tablet drawing app, people made clearer visual elements, sometimes even writing short phrases, transforming the system into shared notes. However, we do not expect users to read messages under their coffee cups. Interestingly, the constraints of drawing functions spurred creativity. For examples, some used the cup stain as a face to add face expressions; some linked all the cups to show them as a group; others encircled their cups with decorative lines to stand out. By providing free form but limited input, self-creation as lightweight interactions opens the creative space for users while avoiding explicit expression or burdensome input.

Comparing to drawing on clay or tablet, water stains faded over time, signifying a transient interaction. Some interactions are inherently transient, such as audio-based interactions seen in the 'Avatar Island' concept. However, visualizations are usually perceived as long-lasting. This prototyping exercise with water stains informed our decision to design transient features explicitly into the experience, ensuring that users understand their temporary nature.

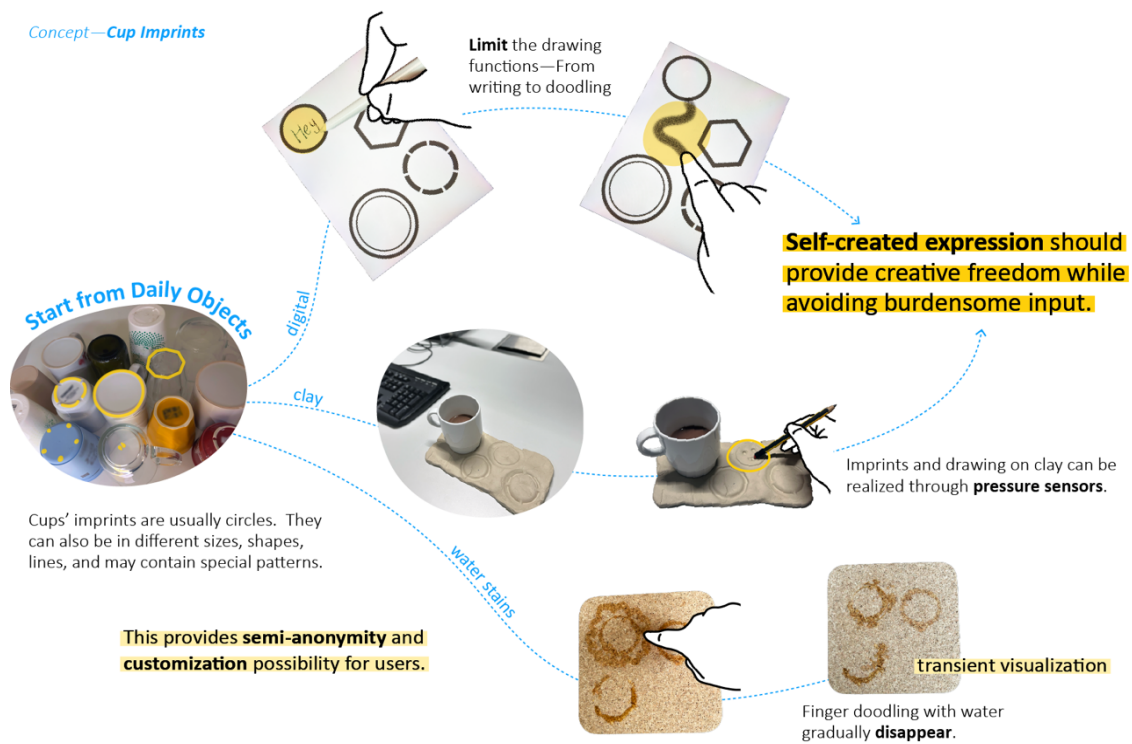

*Figure 8. A visual documentation of prototyping the concept Cup Imprints*

## 5.3  Mapping for versatility

We used LEGO modules to create a miniature office with minifigures representing coworkers. Though initially a visual mock-up for testing sound presets, people instinctively fiddled with the figures, suggesting uses such as moving their locations to indicate activities. Many wanted to customize their own figures and recognize teammates. This installation-art-like design suits small groups with strong bonds due to limited user capacity.

We selected online audio resources addressing the three reasons for using lightweight interactions. To signal presence, we aimed for a sound that subtly surfaces from the background. To exchange responses, we opted for a sound that shows reaction in a brief and firm way. To express intentions, we sought a sound that attracts attention with an engaging tone. After trials, we developed two preset groups: Metaphor sounds from animated movies (e.g., bubbling for 'surfacing up') and human activity sounds (e.g., finger snaps as a natural way to invite attention). These echo non-verbal, face-to-face cues, which are rich in social meanings and can be interpreted in context. In designing presets, we can use daily social practices as design resources to map specific options to versatile meanings.

While experiencing with sound presets, people found human activity sounds intuitive but irritating with repetition. In contrast, metaphor sounds were perceived similarly to auditory feedback in digital systems—learnable and tolerable over time. Additionally, we excluded human voices because participants found voices from a miniature to be intrusive and unnatural. This insight highlights the importance of incorporating abstraction into lightweight interactions to avoid unrealistic expectations from overly human-like representations.

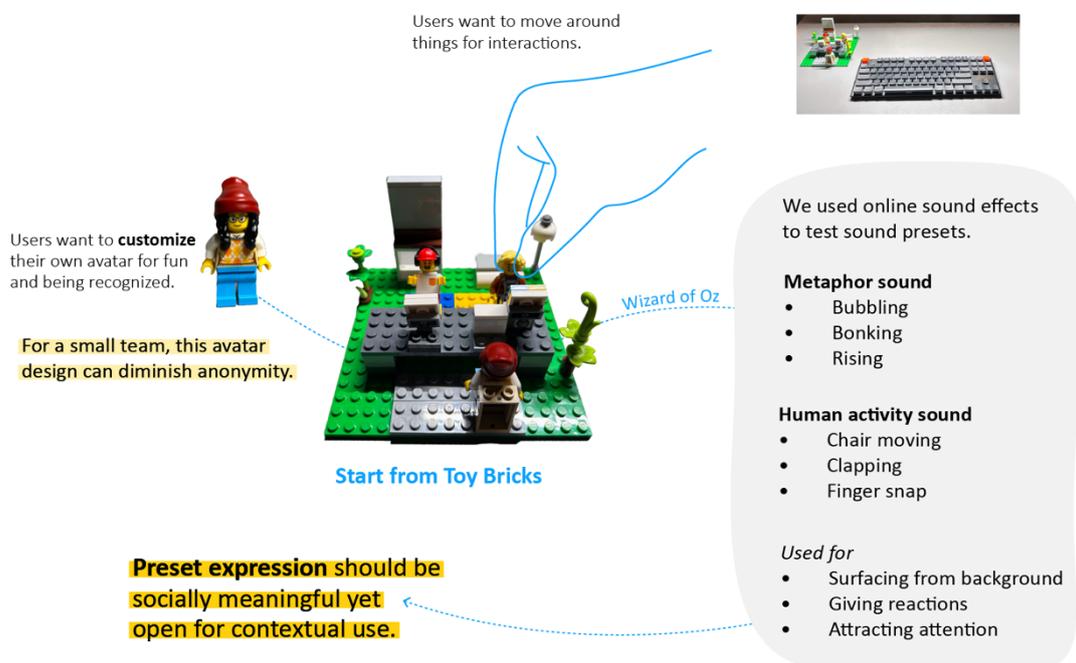

*Figure 9. A visual documentation of prototyping the concept Avatar Island*

# 6 Discussion

## 6.1 Lightweight interactions as catalysts for informal communication

Early HCI and CSCW research framed lightweight interactions as replacements of face-to-face informal conversations on computers, laying the groundwork for today's communication technology (Tang et al., 1994; Whittaker et al., 1997). As the concept of computers evolved, lightweight interactions were referred to quick, simple, and low-effort user-system interactions (Brewer et al., 2007; Gallacher et al., 2015). The designs in this pictorial adopted the latter understanding to enable users to send simple responses to each other. We framed lightweight interactions in awareness-supporting technologies with dual purposes: a design approach aimed for intuitive user input and output; a communication strategy allowing users to express their intentions or respond to others in a natural and simple way.

In the RtD process, we identified three approaches for designing lightweight interactions: using simple expression for intuitive perception, enabling self-created expression for adaptive and personalized usage, and providing preset expression for enriched social practices. With these interactions, our designs aimed to let hybrid workers feel others' social presence, capture the moment of brief social exchanges, and express their communicational intentions in the moment. These implicit yet socially rich moments are not a replacement for but catalysts to informal communication. While developed for hybrid work, the design space for lightweight interactions may extend to other domains requiring social awareness, such as family connections or educational settings, though further research would be needed to validate its application.

## 6.2 Embracing ambiguity in awareness-supporting technologies

In our speculating and prototyping process, the theme of ambiguity emerged as a recurring and valuable design resource. The ambiguity in our designs is threefold:

First, work status information is more of an estimation than an accurate depiction. The status data was collected primarily by detecting one indicative parameter (e.g., pressure on chairs or motion of cups). Explicit manual input or comprehensive multi-parameter detection can be more accurate but intrusive. We argue that an estimation is enough for social awareness because of the inherent ambiguity in the real world (Gaver et al., 2003). For instance, in an office, we gauge overall presence by fragmented information from surroundings.

Second, identity in our designs is (semi-)anonymous, represented through creative proxies (lights, cup imprints, or minifigures) instead of names or photos. Identity information is shaped by user customization, interactions, and temporality, navigating between blending in and standing out within the team. This ambiguity in identity reduces social stress of personal revealing and empowers users with autonomy over their self-presentation, fostering inclusivity (Nissenbaum, 2009; Turkle, 2011).

Third, users need to make sense of lightweight interactions in context. Our lightweight interaction designs encourage spontaneous expression and intrigue users with open interpretations, fostering curiosity and informality within teams. However, this ambiguity also carries the risks of misinterpretation, potentially leading to reduced adoption. Future work should explore how to balance interpretive flexibility with sufficient clarity to minimize misunderstandings and ensure effective communication.

### 6.3 Annotation as tools for analysis and documentation in design research

In this research, we used annotations to bridge design instances and theories. More than documenting outcomes, our annotations functioned as what Schön (1992) calls 'reflection-in-action:' the annotated design knowledge prompted next design moves. In brainstorming, annotations helped surface patterns across disparate ideas. In speculation, we annotated and developed a design space for lightweight interactions, serving as a generative tool. In prototyping, we annotated the interplay between material constraints and experiential qualities, capturing insights emerging from making.

In practice, the annotating process was iterative, dynamic, and often messy. In this pictorial, we aimed to clarify a structured procedure by adopting the layered annotation technique (Vasconcelos et al., 2024). However, it is challenging to reflect the complex process of balancing subjective interpretations with theoretical support. In speculating, we frequently shifted between abstract thinking and concrete composition of design instances. In prototyping, we juggled material exploration, design rationale, and rapid testing. These challenges underscored the non-linear and emergent nature of research-through-design (Auger, 2013; Cross, 1982; Zimmerman et al., 2007). We used this pictorial as an attempt of using annotation to document and communicate a research-through-design process.

## 7 Conclusion

This research explored the design of lightweight interactions in awareness-supporting information technologies through a research-through-design process, emphasizing scenario-based speculation and experience-oriented prototyping. In this annotated documentation, we illuminated the role of lightweight interactions in promoting informal communication and highlighted their design implications for creating inclusive, adaptable, and engaging systems. Future studies can explore how teams appropriate these technologies through sustained use.